\documentclass[aps,pre,groupedaddress]{revtex4-2}

\usepackage[pdftex]{graphicx} 
\usepackage{xcolor} 
\usepackage{mathrsfs}
\usepackage{amsmath,amssymb}
\usepackage{hhline}
\usepackage{dcolumn}
\usepackage{bm}
\usepackage{here}
\usepackage{comment}

\newcommand{\be}{\begin{equation}}
\newcommand{\ee}{\end{equation}}
\newcommand{\ba}{\begin{eqnarray}}
\newcommand{\ea}{\end{eqnarray}}
\newcommand{\bi}{\begin{itemize}}
\newcommand{\ei}{\end{itemize}}

\newcommand{\srt}{s^{\rm rt}}

\newcommand{\bt}{\bar{t}}
\newcommand{\bw}{\bar{w}}
\newcommand{\bv}{\bar{v}}
\newcommand{\btt}{\bar{t}^{(2)}}
\newcommand{\btw}{\bar{w}^{(2,0)}}
\newcommand{\bsw}{\bar{w}^{(1,1)}}
\newcommand{\btv}{\bar{v}^{(2)}}
\newcommand{\bsrt}{\bar{s}^{\rm rt}}

\newcommand{\upp}{U^{\rm oo}}
\newcommand{\upq}{U^{\rm ou}}
\newcommand{\uqq}{U^{\rm uu}}

\newcommand{\bupp}{\bar{u}^{\rm oo}}
\newcommand{\bupq}{\bar{u}^{\rm ou}}
\newcommand{\buqq}{\bar{u}^{\rm uu}}

\newcommand{\buupp}{\bar{u}^{{\rm oo}(2)}}
\newcommand{\buupq}{\bar{u}^{{\rm ou}(2)}}
\newcommand{\buuqq}{\bar{u}^{{\rm uu}(2)}}

\newcommand{\ns}{\mathcal{P}}
\newcommand{\Nns}{\mathcal{N}_n(s)}

\newcommand{\pc}{p_{\rm c}}

\begin{document}
\title{Berezinskii--Kosterlitz--Thouless-type Transition in Site Percolation on the Diamond Hierarchical Lattice}
\author{Takehisa Hasegawa}
\email{takehisa.hasegawa.sci@vc.ibaraki.ac.jp}
\affiliation{Graduate School of Science and Engineering, Ibaraki University, 2-1-1, Bunkyo, Mito, 310-8512, Japan}
\author{Kazuki Wataya}
\affiliation{Graduate School of Science and Engineering, Ibaraki University, 2-1-1, Bunkyo, Mito, 310-8512, Japan}
\author{Tomoaki Nogawa}
\email{nogawa@med.toho-u.ac.jp}
\affiliation{Faculty of Medicine, Toho University, 5-21-16, Omori-Nishi, Ota-ku, Tokyo, 143-8540, Japan}

\begin{abstract}
We study site percolation on the diamond hierarchical lattice, a finite-dimensional fractal network, using an exact generating-function analysis. 
In contrast to bond percolation, site percolation on this lattice does not undergo a transition from a nonpercolating phase to a percolating phase.
Instead, the system exhibits a nonpercolating phase for $p<\pc$ and a critical phase for $p>\pc$. 
In the critical phase, the size of the largest cluster remains subextensive, scaling as $N^{\psi(p)}$, where the fractal exponent $\psi(p)$ varies continuously with $p$.
By analyzing the renormalization-group recursion relation in the vicinity of $\pc$, we show that the correlation length exhibits a Berezinskii--Kosterlitz--Thouless-type essential singularity, $\xi(p)\sim \exp \left({\rm const}/\sqrt{\pc-p}\right)$ for $p \to \pc^-$, which is further confirmed by finite-size scaling analyses showing excellent data collapse. 
These results demonstrate that critical phases in percolation can emerge even on finite-dimensional networks and that exponential volume growth is not necessary for such phases to appear.
We argue that the critical phase on the diamond hierarchical lattice stems from site dilution remaining relevant under renormalization. 
\end{abstract}

\maketitle

\section{Introduction}

Percolation theory~\cite{stauffer2018introduction,li2021percolation} provides a basic paradigm for connectivity-driven phase transitions in lattices and complex networks~\cite{newman2018networks, dorogovtsev2022nature}.
As the occupation probability of nodes or, alternatively, the open probability of edges, denoted by $p$, increases, a system typically undergoes a second-order transition from a nonpercolating phase in which only finite clusters exist to a percolating phase characterized by the emergence of a giant cluster occupying a finite fraction of the system.
The critical behavior near the transition point is characterized by critical exponents and scaling relations~\cite{li2021percolation}.

The two fundamental percolation processes are bond percolation and site percolation.
On Euclidean lattices, bond and site percolation transitions generally share the same overall phase-diagram structure. 
Moreover, they are believed to belong to the same universality class, with identical critical exponents~\cite{stauffer2018introduction}.
However, this equivalence is not guaranteed on inhomogeneous networks.
For example, Radicchi and Castellano~\cite{radicchi2015breaking} showed that on a heavy-tailed scale-free network the critical exponent $\beta$ of the order parameter differs by unity between bond and site percolation.

An even more striking difference between bond and site percolation occurs in the Dorogovtsev--Goltsev--Mendes network~\cite{dorogovtsev2002pseudofractal}, also known as the $(1,2)$-flower~\cite{rozenfeld2007percolation}.
On this network, the phase diagrams of bond and site percolation are completely different.
For bond percolation, the system remains in the percolating phase for any $p>0$~\cite{dorogovtsev2003renormalization}.
In contrast, for site percolation the system is in a {\it critical phase}~\cite{nogawa2009monte,hasegawa2013profile,hasegawa2014critical} 
(also known as the patchy phase~\cite{boettcher2009patchy,boettcher2015classification}) for any $0<p<1$.
This phase is characterized by a power-law cluster-size distribution with a continuously varying exponent $\tau(p)$, although such power-law behavior is typically observed only at a second-order transition point.

Hierarchical networks provide a useful framework for studying unconventional percolation phenomena.
Their recursive construction allows exact renormalization-group (RG) transformations to be implemented, enabling critical points and critical behaviors to be determined analytically.
Critical phases have been identified in RG analyses of bond percolation on hierarchical networks with the small-world property~\cite{boettcher2009patchy,hasegawa2010generating,boettcher2012ordinary,nogawa2014transition,nogawa2018renormalization}.
In particular, bond percolation on the decorated $(2,2)$-flower~\cite{rozenfeld2007percolation} exhibits both a critical phase and a percolating phase~\cite{berker2009critical,hasegawa2010generating}. 
This network is obtained by adding shortcut edges to the finite-dimensional diamond hierarchical lattice (DHL) [Fig.~\ref{fig:construction}(b)], also known as the $(2,2)$-flower, which induces the small-world property.
From the RG viewpoint, a critical phase is associated with nontrivial stable fixed points.
Moreover, the transition from the critical phase to the percolating phase is accompanied by an essential singularity in the order parameter, often referred to as an inverted Berezinskii--Kosterlitz--Thouless (BKT) transition~\cite{berker2009critical}.

In studies of nonamenable graphs, such as hyperbolic lattices and trees, a critical phase has been characterized as a region in which the correlation length remains finite while the correlation volume, defined as the sum of two-point correlations between a given node and all other nodes, diverges~\cite{nogawa2009monte,baek2009comment,nogawa2009reply,hasegawa2014critical}.
In other words, although the probability that a given node belongs to the same cluster as any particular distant node decays exponentially with distance, the cluster containing that node can still extend to arbitrarily large distances.
This scenario requires the number of reachable nodes to grow exponentially with the distance from a given node.
Consistent with this view, critical phases have been identified not only in nonamenable graphs~\cite{benjamini1996percolation,benjamini2001percolation,nogawa2009monte,baek2009comment,nogawa2009reply,gu2012crossing}, but also in growing networks with the small-world property~\cite{hasegawa2010critical,hasegawa2013profile} and in hierarchical small-world networks~\cite{boettcher2009patchy,boettcher2012ordinary,hasegawa2013absence,singh2014scaling,nogawa2014transition,wataya2025critical}.
These observations raise the following question: Is exponential volume growth a necessary condition for the emergence of critical phases in percolation, or can hierarchical structure alone give rise to such behavior?

In this paper, we address this question by studying site percolation on the DHL, a hierarchical network with fractal dimension $d_{\rm f}=2$~\cite{yamamoto2023bifractality}.
Previous work showed that, in the decorated $(2,2)$-flower, an arbitrarily small amount of site dilution (node removal) eliminates the percolating phase, irrespective of the fraction of shortcut edges~\cite{hasegawa2012phase}. 
This observation suggests that site percolation on the DHL may exhibit behavior fundamentally distinct from that of bond percolation, even in finite-dimensional systems.

Motivated by this possibility, we perform a detailed generating-function analysis of site percolation on the DHL.
We show that the system exhibits a nonpercolating phase and a critical phase, but no percolating phase.
The critical phase identified here emerges without exponential volume growth, in sharp contrast to previously known examples.
The correlation length remains infinite throughout the critical phase, while no giant cluster ever forms.
We further demonstrate that the transition is governed by a BKT-type essential singularity analogous to that in the two-dimensional XY model~\cite{berezinskii1971destruction,kosterlitz1973ordering}.
However, the underlying mechanism is specific to hierarchical systems, where the effects of site dilution persist under coarse graining at every level.
Our results reveal a distinct route to the emergence of critical phases in finite-dimensional hierarchical networks and highlight a fundamental distinction between site and bond percolation rooted in hierarchical structure.

\section{Diamond Hierarchical Lattice and Site Percolation}

\begin{figure}[t]
\centering
(a)
\includegraphics[width=0.6\linewidth]{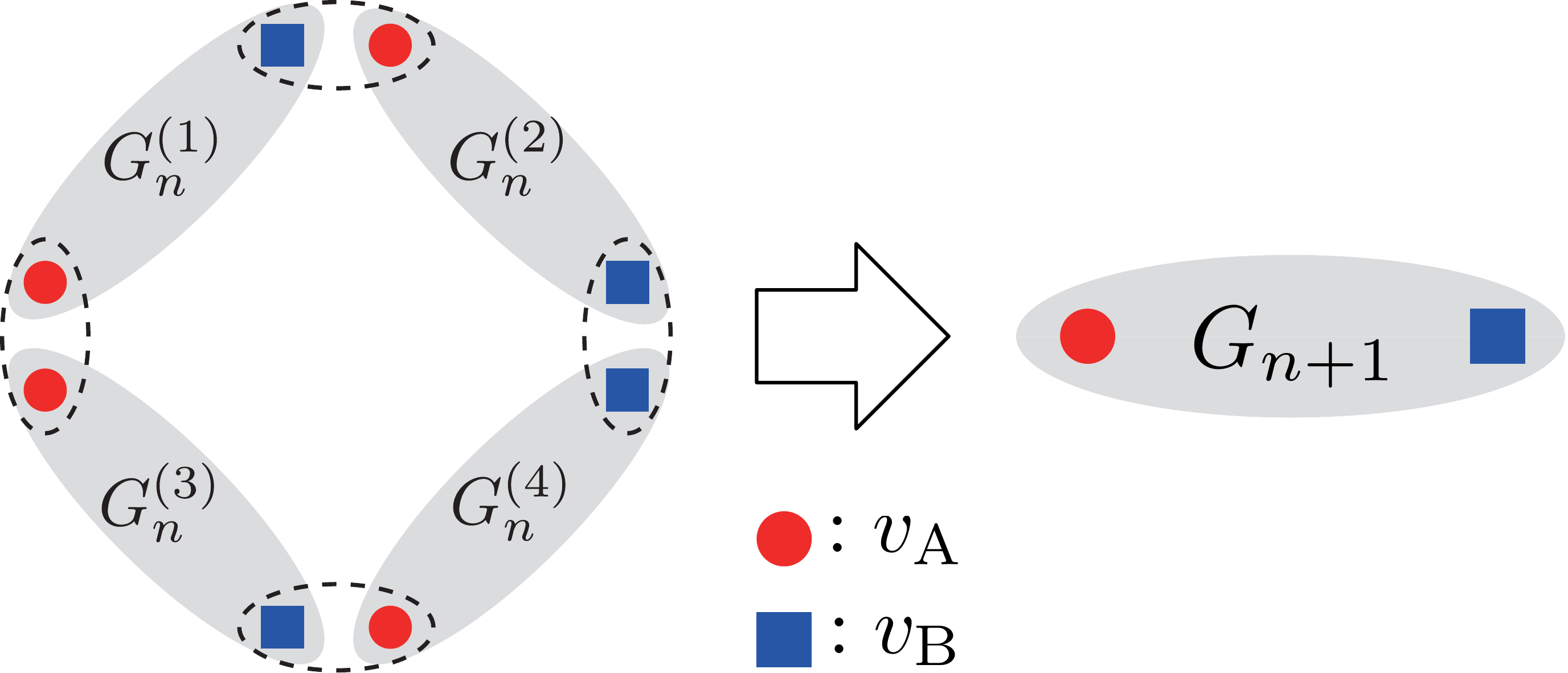}
(b)
\includegraphics[width=0.25\linewidth]{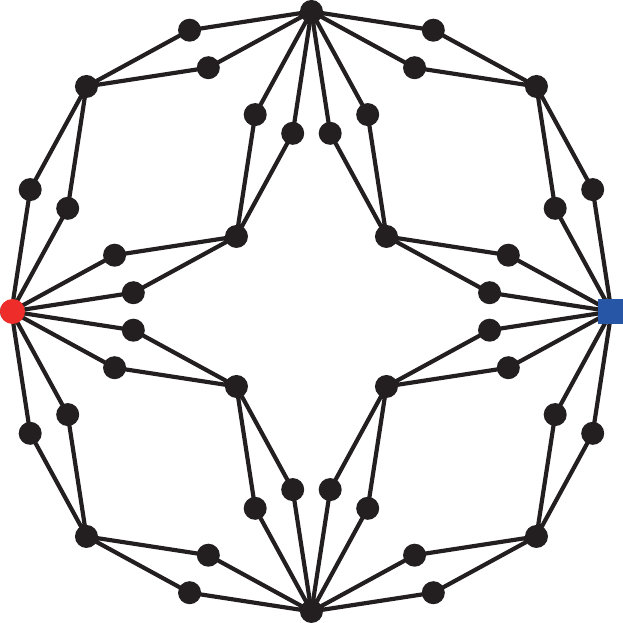}
\caption{
Construction of the DHL. 
(a) The graph $G_{n+1}$ is obtained by merging the roots of four copies of $G_n$ as shown here.
Two of the merged nodes become the roots of $G_{n+1}$, denoted by $v_{\rm A}$ and $v_{\rm B}$. 
(b) Example of the fourth-generation graph $G_4$.
}
\label{fig:construction}
\end{figure}

This study focuses on the DHL.
This graph is generated recursively by applying a deterministic replacement rule, as illustrated in Fig.~\ref{fig:construction}(a).
Let $G_n$ denote the graph at generation $n$.
At $n=1$, $G_1$ consists of two nodes connected by a single edge; we call these nodes the {\it roots}, $v_{\rm A}$ and $v_{\rm B}$.
The graph $G_{n+1}$ can be constructed by merging four copies of $G_n$.
Let $G_n^{(i)}$ ($i=1,2,3,4$) be copies of $G_n$, whose roots are denoted by $v_{\rm A}^{(i)}$ and $v_{\rm B}^{(i)}$.
Then $G_{n+1}$ is obtained by the following identifications [Fig.~\ref{fig:construction}(a)]:
(i) merge $v_{\rm B}^{(1)}$ with $v_{\rm A}^{(2)}$;
(ii) merge $v_{\rm B}^{(3)}$ with $v_{\rm A}^{(4)}$;
(iii) merge $v_{\rm A}^{(1)}$ with $v_{\rm A}^{(3)}$, and denote the resulting node by the next-generation root $v_{\rm A}$;
(iv) merge $v_{\rm B}^{(2)}$ with $v_{\rm B}^{(4)}$, and denote the resulting node by the next-generation root $v_{\rm B}$.
Figure~\ref{fig:construction}(b) shows the fourth-generation graph $G_4$.

The numbers of nodes and edges of $G_n$ are $N_n=\tfrac{2}{3}(4^{n-1}+2)$ and $M_n=4^{\,n-1}$, respectively.
At $n=1$, the graph consists of two nodes of degree $1$.
For $n \ge 2$, replication increases the number of degree-$k$ nodes fourfold at each generation.
In addition, the pairwise merging of the copied roots produces four nodes whose degree is twice the largest degree in the previous generation.
Accordingly, the number of nodes with degree $k_m=2^m$ ($m=1,2,\dots,n-1$) is given by $N_n(k_m) = 2\cdot4^{\,n-1-m}$ for $m<n-1$, and $N_n(k_m)=4$ for $m=n-1$.
These relations yield the asymptotic degree distribution $P(k)\propto k^{-3}$ for $n\gg1$. 
The roots $v_{\rm A}$ and $v_{\rm B}$ have the largest degree $k_{\rm max}=2^{n-1}$.
Moreover, the lattice is a fractal with fractal dimension $d_{\rm f}=2$~\cite{rozenfeld2007percolation} (a more refined characterization reveals bifractal properties~\cite{yamamoto2023bifractality}).
Let $L_n$ be the shortest-path distance between the two roots in $G_n$.
By construction,
\be
L_n=2^{\,n-1}\propto N_n^{1/2},
\ee
indicating that the DHL is effectively two-dimensional in terms of the relation between linear size and system size.

We analyze the percolation problem on the DHL by focusing on the connectivity between the two roots.
Before discussing site percolation, we briefly review bond percolation on the DHL for comparison, in which each edge is open with probability $p$ and closed otherwise.
We define the contact probability $P_n^{\rm (bond)}$ as the probability that the two roots of $G_n$ are connected by open edges.
Owing to the hierarchical structure of the DHL, $P_n^{\rm (bond)}$ satisfies the exact recursion relation $P_{n+1}^{\rm (bond)}=1-\bigl[1-\bigl(P_n^{\rm (bond)}\bigr)^2\bigr]^2$, with the initial condition $P_1^{\rm (bond)}=p$.
This equation exhibits a conventional transition from a nonpercolating phase to a percolating phase at bond-percolation critical point $\pc^{\rm (bond)}=(\sqrt{5}-1)/2$.
The RG analysis also yields critical exponents that do not coincide with those of two-dimensional systems~\cite{rozenfeld2007percolation}.

The main focus of the present work is site percolation.
Each node is occupied with probability $p$ and unoccupied with probability $q=1-p$.
We define $P_n$ as the probability that the two roots are connected via occupied nodes, {\it conditional on both roots being occupied}.
Then the unconditional contact probability $P_n^{\rm (site)}$ is
\be
P_n^{\rm (site)} = p^2 P_n.
\ee
The recursive structure leads to the exact recursion relation for the conditional probability $P_n$ as
\be
1-P_{n+1}=[q + p (1-P_n^2)]^2,
\ee
or equivalently,
\be
P_{n+1}=R(P_n; p)=2pP_n^2-p^2P_n^4,
\label{eq:recursion_Pn_site}
\ee
with the initial condition $P_1=1$.
Fixed points $P_\ast$ satisfying $R(P_\ast; p)=P_\ast$ characterize the asymptotic behavior of $P_n$ as $n\to\infty$.
In the site-percolation case, the recursion map $R(P; p)$ explicitly depends on the occupation probability $p$, and therefore the fixed-point structure changes with $p$.
We define the site-percolation critical point $\pc$ as the point at which the nontrivial stable and unstable fixed points merge and become marginal:
\be
\left.\frac{dR}{dP}\right|_{P=P_\ast}=1.
\ee
This behavior corresponds to a saddle-node bifurcation of the recursion map.
Solving these equations, we obtain
\be
\pc=\frac{27}{32},\qquad P_\ast(\pc)=\frac{8}{9}. \label{eq:criticalPoint}
\ee
For $p<\pc$, the iteration converges to $P_n\to 0$, while for $p \ge \pc$ it converges to $P_n\to P_\ast>0$.

In the next section, we introduce generating functions to evaluate the mean size of the root cluster, the cluster size distribution, and the susceptibility, defined as the mean size of clusters excluding those containing the roots.
This formulation enables an exact analysis of the phase transition and critical phenomena in site percolation on the DHL.

\section{Generating Function Method}\label{sec:generating}

In this section, we apply a generating-function method to analyze site percolation on the DHL.
Our primary focus is on clusters containing at least one of the two roots, $v_{\rm A}$ and $v_{\rm B}$, which we collectively refer to as {\it root clusters}.
The scaling behavior of the root-cluster size serves to distinguish the nonpercolating and critical phases in the absence of a giant cluster.

To characterize the statistics of root clusters in $G_n$, we introduce the following three probability distributions.
(i) Let $t_n(s)$ be the probability that, conditioned on both roots being occupied, they are connected and the cluster containing both roots has size $s$.
(ii) Let $w_n(s_1,s_2)$ be the probability that, conditioned on both roots being occupied, the two roots are not connected and the clusters containing the two roots have sizes $s_1$ and $s_2$, respectively.
(iii) Let $v_n(s)$ be the probability that, conditioned on exactly one of the two roots being occupied, the cluster containing the occupied root has size $s$.
In all cases, cluster sizes are defined to exclude the roots themselves.

We introduce the generating functions for $t_n(s)$, $w_n(s_1,s_2)$, and $v_n(s)$ as
\ba
T_n(x) &=& \sum_{s} t_n(s) x^s, \label{eq:defT} \\
W_n(x,y) &=& \sum_{s_1}\sum_{s_2} w_n(s_1, s_2) x^{s_1} y^{s_2}, \label{eq:defW}\\
V_n(x) &=& \sum_{s} v_n(s) x^s. \label{eq:defV}
\ea
By definition, these functions satisfy
\be
T_n(1) = P_n , \quad
W_n(1,1) = Q_n  \equiv 1 - P_n, \quad
V_n(1) = 1.
\ee

Because $G_{n+1}$ is constructed by merging four copies of $G_n$ and the site percolation configurations on these copies are independent, we obtain the following recursion relations for the generating functions:
\ba
T_{n+1}(x) 
&=& p^2 [x^2 T_n^4(x) + 4 x^2 T_n^3(x) W_n(x,x) + 2x T_n^2(x) W_n^2(x,1)] + 2 p q x T_n^2(x) V_n^2(x), \label{eq:Trec} \\
W_{n+1}(x,y) 
&=& p^2 [ x T_n(x) W_n(x,y) + y T_n(y) W_n(x,y) + W_n(x,1) W_n(y,1)]^2 \nonumber \\
&&+ 2pq V_n(x) V_n(y) [x T_n(x) W_n(x,y) + y T_n(y) W_n(x,y) + W_n(x,1) W_n(y,1)] \label{eq:Wrec} \\
&&+q^2 V_n^2(x) V_n^2(y), \nonumber \\
V_{n+1}(x) &=& p^2 [W_n(x,1) + xT_n(x)V_n(x)]^2
+2pqV_n(x)[W_n(x,1) + xT_n(x)V_n(x)] +q^2 V_n^2(x), \label{eq:Vrec}
\ea
with the initial conditions
\be
T_1(x) =  V_1(x) =1, \qquad W_1(x,y) = 0.
\ee
Each term in Eqs.~(\ref{eq:Trec})--(\ref{eq:Vrec}) corresponds to a distinct connectivity pattern among the four subgraphs ($G_n^{(1)}, \ldots, G_n^{(4)}$) constituting $G_{n+1}$. 
The factors involving $p$ and $q$ reflect the occupation states of the two internal roots formed by merging the roots of the constituent copies of $G_n$, and the numerical coefficients account for combinatorial multiplicities.

Let $\srt_n$ denote the size of the cluster containing the root $v_{\rm A}$ in $G_n$.
The generating function for its size distribution is given by
\begin{equation}
\sum_s {\rm Pr}(\srt_n = s) x^s = q+px [pxT_n(x)+pW_n(x,1)+qV_n(x)]. \label{eq:rootGF}
\end{equation}
The mean size $\bsrt_n(p)$ is then given by
\be
\bsrt_n(p) = \frac{d}{dx} \sum_s {\rm Pr}(\srt_n = s) x^s \Big|_{x=1} = p (1+ p P_n + p \bt_n + p\bw_n+q\bv_n),
\label{eq:sroot}
\ee
where
\be
\bt_n = \frac{d}{dx}T_n(x)\Big|_{x=1}, \quad
\bw_n = \frac{d}{dx}W_n(x,1)\Big|_{x=1}= \frac{d}{dx}W_n(1,x)\Big|_{x=1}= \frac{1}{2}\frac{d}{dx}W_n(x,x)\Big|_{x=1}, \quad
\bv_n = \frac{d}{dx}V_n(x)\Big|_{x=1}.
\ee
The quantities $\bar{t}_n$, $\bar{w}_n$, and $\bar{v}_n$ are evaluated recursively from the derivatives of Eqs.~(\ref{eq:Trec})--(\ref{eq:Vrec}); their explicit recursion relations are given in Appendix~\ref{sec:recForms}.
The normalized mean root-cluster size, $\bsrt_n(p)/N_n$, plays the role of an order parameter; it remains finite when a giant cluster exists, whereas it vanishes otherwise.

The fractal exponent $\psi(p)$ characterizes the scaling behavior of the mean root-cluster size as
\be
\bsrt_n(p)\propto N_n^{\psi(p)}
\qquad (n\gg1)
\ee
and provides a useful indicator for distinguishing between the nonpercolating and critical phases~\cite{nogawa2009monte}.
Using $\bsrt_n(p)$, we define a finite-size estimate of the fractal exponent, $\psi_n(p)$, for $G_n$ as
\be
\psi_n(p) = \frac{\ln \bsrt_{n+1}(p) - \ln \bsrt_{n-1}(p)}{\ln N_{n+1} - \ln N_{n-1}} \approx \frac{{\rm d} \ln \bsrt_n(p)}{{\rm d}\ln N_n}.
\ee
The fractal exponent is then defined by the limit
\be
\psi(p) = \lim_{n \to \infty} \psi_n(p).
\ee
In degree-bounded networks, $\psi(p)=0$ corresponds to a nonpercolating phase, while a critical phase is characterized by a continuously varying exponent satisfying $0<\psi(p)<1$.
A percolating phase, when present, corresponds to $\psi(p)=1$.
However, for inhomogeneous networks in which the largest degree scales as $k_{\rm max} \sim O(N^{\alpha})$, the fractal exponent can take a $p$-independent nonzero value, $\psi(p)=\alpha$, even though long-range connectivity is absent (i.e., the system remains in the nonpercolating phase), as demonstrated for the Song--Havlin--Makse network~\cite{wataya2025critical}.
In such cases, a critical phase is characterized by a continuously varying exponent satisfying $\alpha<\psi(p)<1$.

Furthermore, the cluster-size distribution and the susceptibility can also be evaluated using generating functions.
Let $\Nns$ denote the number of clusters of size $s$ in $G_n$, excluding the root clusters.
The corresponding cluster-size distribution $\ns_n(s)$ is defined as
\be
\ns_n(s) = \frac{\Nns}{N_n}.
\ee
Introducing the generating function
\be
U_n(x)  = \sum_s \Nns x^s, \label{eq:defU}
\ee
we can express $\ns_n(s)$ in terms of $U_n(x)$ as
\be
\ns_n(s) = \frac{1}{N_n} \frac{1}{s!}\frac{d^s}{dx^s} U_n(x) \Big|_{x=0}. \label{eq:nsFromUn}
\ee
The susceptibility $\chi_n(p)$ is also obtained from $U_n(x)$ as
\be
\chi_n(p) = \sum_s s^2 \ns_n(s) = \frac{1}{N_n} [U_n''(1) + U_n'(1)],
\ee
where $U_n'(1)$ and $U_n''(1)$ are evaluated recursively.
Details of the calculation of $U_n(x)$ and $\chi_n(p)$ are presented in Appendix~\ref{sec:recForms}.

\begin{figure}[t]
\centering
(a)
\includegraphics[width=0.45\linewidth]{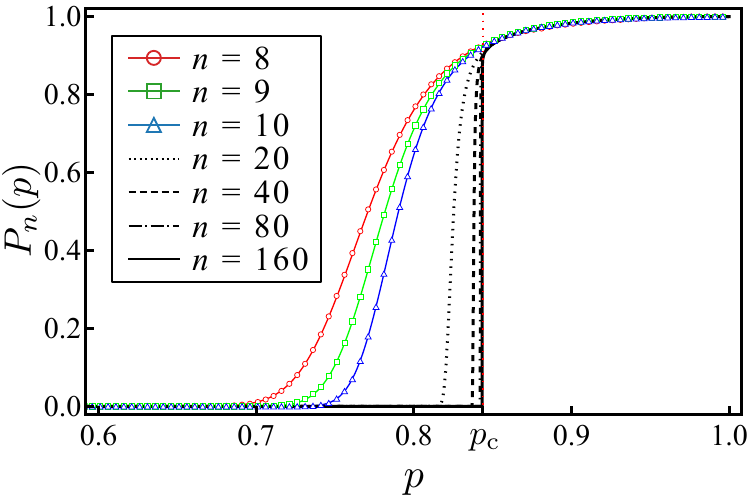}
(b)
\includegraphics[width=0.45\linewidth]{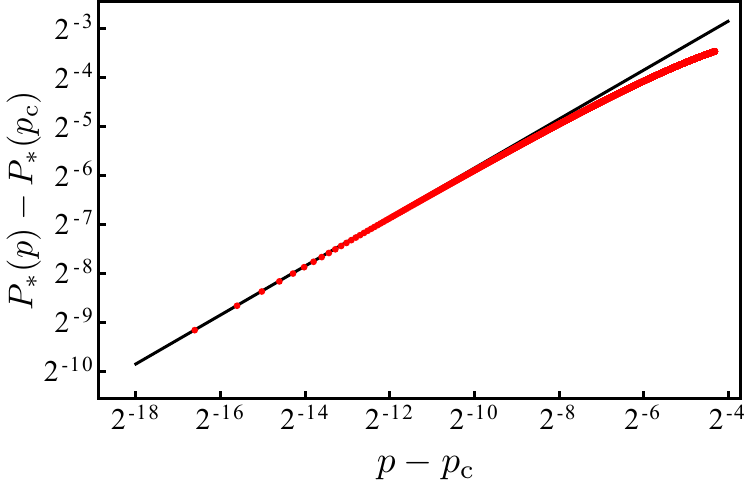}
\caption{
(a) 
Conditional contact probability $P_n$ as a function of the occupation probability $p$ for site percolation on the DHL.
The thin red, green, and blue curves correspond to generations $n=8$, $9$, and $10$, respectively.
The dotted, dashed, dot-dashed, and solid curves correspond to the larger generations $n=20$, $40$, $80$, and $160$.
The lines show analytical results obtained from the recursion relations, while the symbols represent Monte Carlo simulation data.
As the generation $n$ increases, $P_n(p)$ becomes progressively sharper and, in the limit $n \to \infty$, exhibits a finite jump at the critical point $\pc$ (vertical dotted line), followed by a continuous increase for $p>\pc$.
(b) 
Log-log plot of the deviation of the stable fixed point, $P_\ast(p)-P_\ast(\pc)$, as a function of $p-\pc$.
The black line, drawn from Eq.~(\ref{eq:Deltadelta}), has slope $1/2$.
The data exhibit the power-law behavior $P_\ast(p)-P_\ast(\pc)\propto (p-\pc)^{1/2}$, indicating a square-root singularity of the fixed point at $\pc$.
}
\label{fig:percProb}
\end{figure}

\begin{figure}[t]
\centering
(a)
\includegraphics[width=0.45\linewidth]{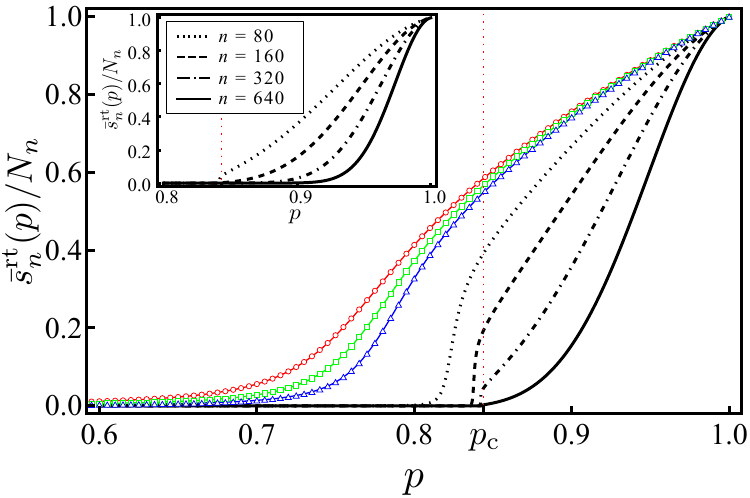}
(b)
\includegraphics[width=0.45\linewidth]{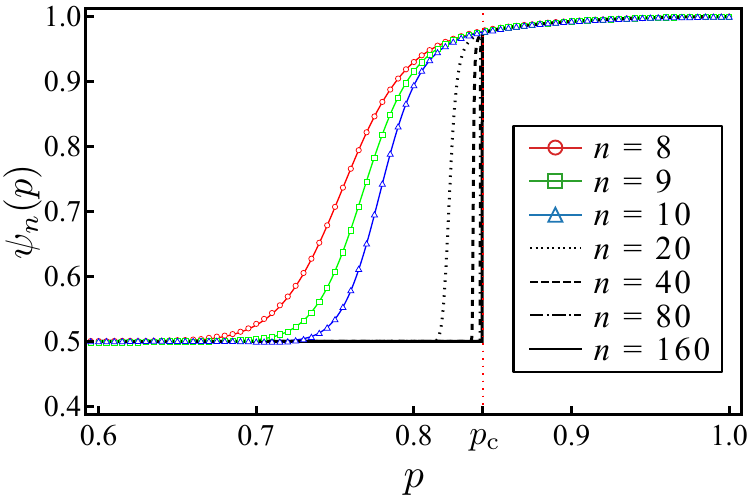}
\caption{
(a) Normalized mean root-cluster size $\bsrt_n(p)/N_n$ and (b) fractal exponent $\psi_n(p)$ as functions of the occupation probability $p$ for site percolation on the DHL.
The thin red, green, and blue curves correspond to generations $n=8$, $9$, and $10$, respectively.
The dotted, dashed, dot-dashed, and solid curves correspond to the larger generations $n=20$, $40$, $80$, and $160$.
The lines show analytical results obtained from the recursion relations, while the symbols represent Monte Carlo simulation data.
In panel (a), the growth of $\bsrt_n(p)/N_n$ becomes increasingly suppressed over the entire range of $p$ as the generation $n$ increases.
Inset of panel (a): normalized mean root-cluster size $\bsrt_n(p)/N_n$ for larger generations $n=80$, $160$, $320$, and $640$ (dotted, dashed, dot-dashed, and solid curves, respectively), highlighting the absence of a giant cluster for any $p<1$.
In panel (b), $\psi_n(p)$ takes the $p$-independent value $1/2$ for $p<\pc$, exhibits a finite jump at the critical point $\pc$ in the large-$n$ limit, and varies continuously with $p$ for $p>\pc$, consistent with the critical phase.
}
\label{fig:sroot}
\end{figure}

\begin{figure}[t]
\centering
(a)
\includegraphics[width=0.5\linewidth]{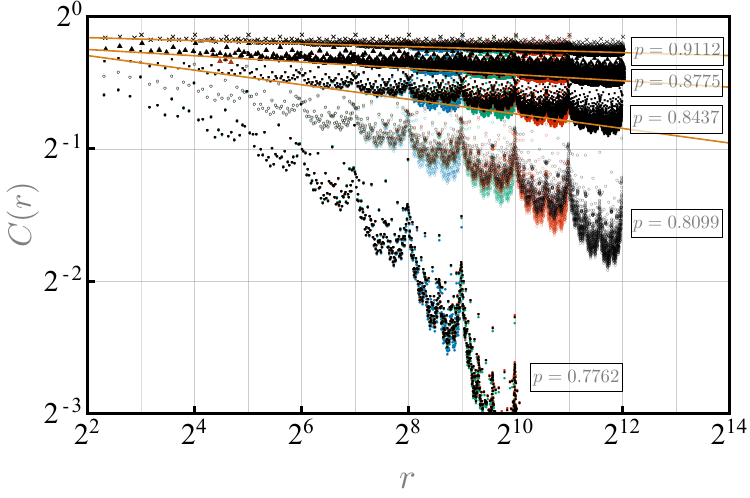}
(b)
\includegraphics[width=0.4\linewidth]{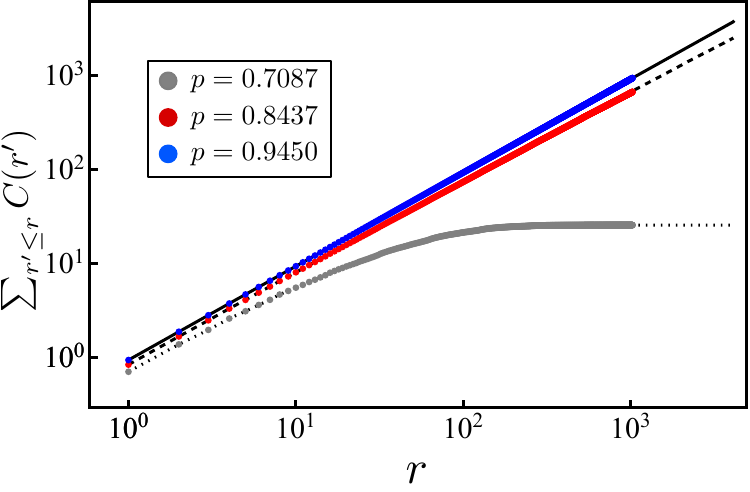}
\caption{
(a) Log-log plot of the correlation function $C(r)$ as a function of the distance $r$ from the root $v_{\rm A}$ for site percolation on the DHL.
Here, $C(r)$ denotes the probability that a node at distance $r$ from the root $v_{\rm A}$ belongs to the root cluster.
Monte Carlo simulation data obtained from $2^{16}$ samples are shown, with different symbols representing different values of $p$.
Blue, green, red, and black symbols correspond to generations $n= 10$, $11$, $12$, and $13$, respectively.
For $p \ge \pc$, the data are consistent with a power-law decay over a broad range of distances.
The solid lines are guides to the eye indicating the form $C(r)\sim r^{-2[1-\psi(p)]}$.
(b) Cumulative correlation function $\sum_{r' \le r} C(r')$ as a function of $r$.
Symbols and curves show Monte Carlo simulation data obtained from $2^{16}$ samples for generations $n=11$ and $13$, respectively.
To make the saturation below $\pc$ more clearly visible, the data for $p < \pc$ are shown for a smaller value of $p$ than in panel (a).
For $p < \pc$, the cumulative correlation function approaches a constant, whereas at $p = \pc$ and for $p > \pc$ it continues to grow with $r$, indicating the absence of a finite correlation length.
}
\label{fig:distance}
\end{figure}

\begin{figure}[t]
\centering
\includegraphics[width=0.45\linewidth]{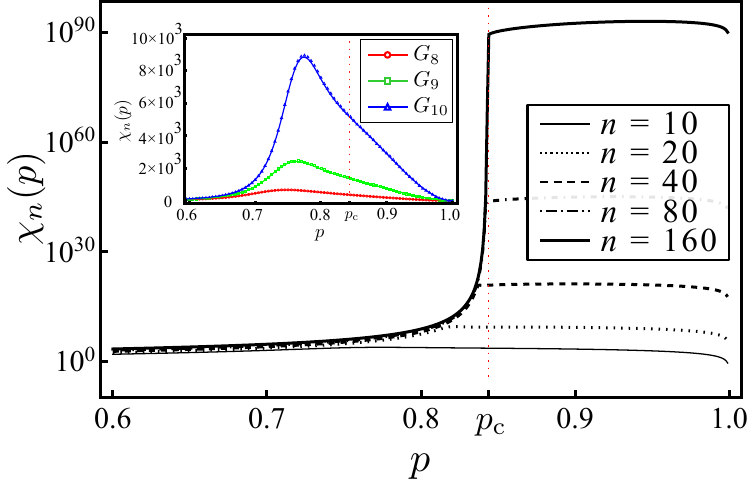}
\caption{
Susceptibility $\chi_n(p)$ as a function of $p$ for site percolation on the DHL.
The thin solid, dotted, dashed, dot-dashed, and thick solid lines correspond to generations $n=10$, $20$, $40$, $80$, and $160$, respectively.
The lines are obtained from the recursion relations.
The susceptibility exhibits a pronounced maximum below the critical point $\pc$.
As the generation $n$ increases, the peak height grows and its position shifts toward $\pc$, while $\chi_n(p)$ remains large throughout the critical phase for $p>\pc$.
Inset: comparison between the results obtained from the recursion relations (lines) and Monte Carlo simulation data (symbols) for smaller generations $n=8$, $9$, and $10$, showing excellent agreement between the two approaches.
}
\label{fig:susceptibility}
\end{figure}

\begin{figure}[t]
\centering
(a)
\includegraphics[width=0.3\linewidth]{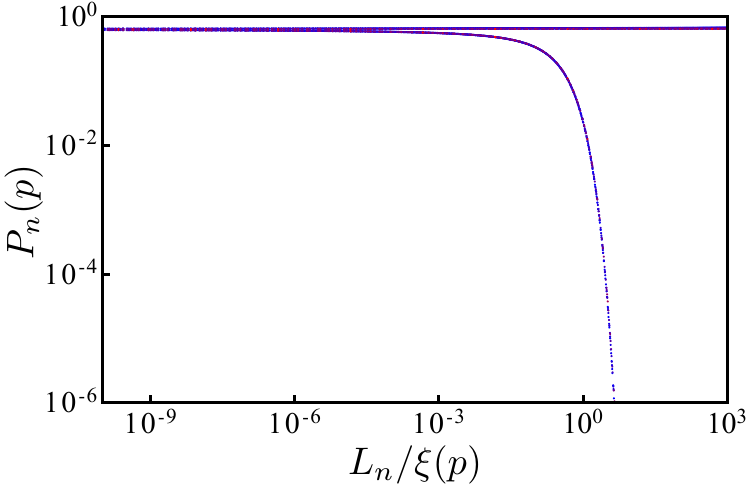}
(b)
\includegraphics[width=0.3\linewidth]{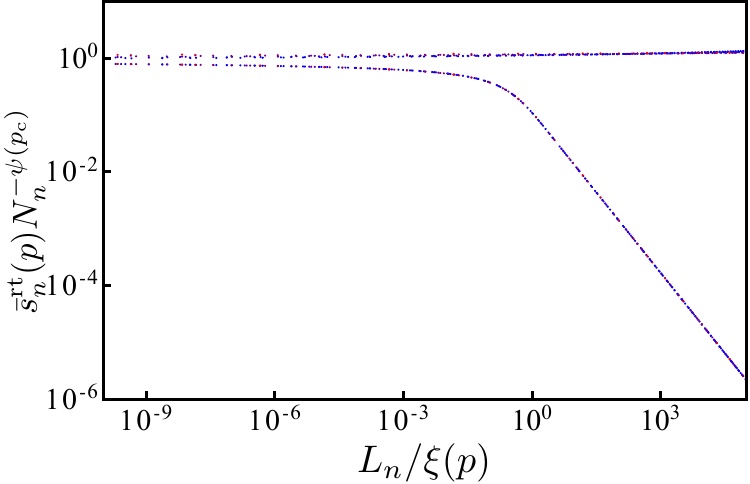}
(c)
\includegraphics[width=0.3\linewidth]{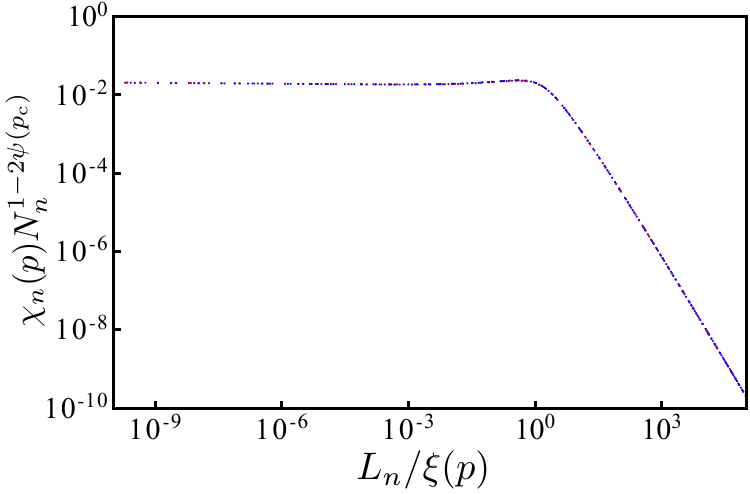}
\caption{
Finite-size scaling plots of (a) the conditional contact probability $P_n(p)$, (b) the mean root-cluster size $\bsrt_n(p)$, and (c) the susceptibility $\chi_n(p)$, as functions of the scaled variable $L_n/\xi(p)$.
Data are shown for generations $n=50$ to $150$ in steps of $5$.
The correlation length $\xi(p)$ is assumed to take the form given in Eq.~\eqref{eq:corrLength}.
All data collapse onto universal curves, providing strong evidence that the correlation length diverges with an essential singularity at the critical point $\pc$.
}
\label{fig:scaling}
\end{figure}

\section{Results}

In this section, we investigate the critical behavior of site percolation on the DHL by evaluating the quantities formulated in Sec.~\ref{sec:generating}.
All analytical results are obtained from exact recursion relations.
As an independent check of the analytical results,  we also carried out Monte Carlo simulations for finite generations, $G_8$, $G_9$, and $G_{10}$, using the Newman--Ziff algorithm~\cite{newman2001fast}, averaging over $10^5$ independent samples.

First, we consider the conditional contact probability $P_n$, defined as the probability that the two roots in $G_n$ are connected, given that both roots are occupied.
Figure~\ref{fig:percProb}(a) shows $P_n$ as a function of the occupation probability $p$ for generations ranging from $n=8$ to $n=160$.
For the smaller generations $n=8, 9$, and $10$, the analytical results obtained from the generating-function method are in excellent agreement with the Monte Carlo results.
These results also show that the convergence of $P_n$ is extremely slow: even for the largest generations accessible to our Monte Carlo simulations, the asymptotic behavior is still far from being reached.
Observing the asymptotic behavior requires generations as large as $n\sim 160$, corresponding to system sizes of order $N_n\sim 10^{96}$.
As $n$ increases, $P_n$ (and hence the contact probability $P_n^{\rm (site)}$) converges to zero for $p<\pc$, while it approaches a finite value $P_\ast>0$ for $p>\pc$, where $\pc=27/32$ as given by Eq.~\eqref{eq:criticalPoint}.

Figure~\ref{fig:percProb}(b) plots $P_\ast(p)-P_\ast(\pc)$ as a function of $p-\pc$ in the region $p>\pc$.
The numerical data confirm the square-root singularity,
\be
P_\ast(p)-P_\ast(\pc) \propto (p-\pc)^{1/2}, \label{eq:squareRoot}
\ee
consistent with the saddle-node nature of the RG fixed point.
In fact, this functional form can also be derived analytically (see Appendix~\ref{sec:square}).

Figure~\ref{fig:sroot}(a) shows the normalized mean size of the root cluster, $\bsrt_n(p)/N_n$, for generations $n=8$ to $n=160$.
For generations accessible to Monte Carlo simulations, $\bsrt_n(p)/N_n$ appears to take nonzero values over a broad range of $p$, which might be interpreted as evidence for a giant cluster.
However, the analytical results for much larger generations reveal that this behavior is purely a finite-size effect.
For $n \gtrsim 80$, the growth of $\bsrt_n(p)/N_n$ becomes increasingly downward-convex, and the quantity decreases over the entire range of $p<1$ as $n$ increases, indicating the absence of a giant cluster on the DHL.

No giant cluster exists for $p<1$.
However, the absence of a giant cluster does not necessarily imply a nonpercolating phase; for $p>\pc$, the system remains in a critical phase.
In this regime, the root-cluster size does not scale extensively with the system size, but instead follows a sublinear scaling form $\bsrt_n(p) \sim N_n^{\psi(p)}$ with $1/2 < \psi(p) < 1$.
Figure~\ref{fig:sroot}(b) shows the finite-size estimate of the fractal exponent, $\psi_n(p)$, as a function of $p$.
Above the critical point, $\psi_n(p)$ takes an approximately size-independent finite value strictly between $1/2$ and $1$ and this value varies continuously with $p$, which is a hallmark of a critical phase.
The cluster-size distribution $\ns_n(s)$ also exhibits a power-law form for $p>\pc$ and the corresponding exponent $\tau(p)$ varies continuously with $p$, reflecting the criticality of this phase (see Appendix~\ref{sec:ns}).

As the occupation probability $p$ is decreased from $p=1$, the fractal exponent $\psi(p)$ decreases accordingly and takes the value $\psi(\pc) \approx 0.972$ at the critical point $\pc$.
For $p<\pc$, the exponent converges in the large-$n$ limit to a $p$-independent value, $\psi(p)=1/2$.
This value coincides with the scaling of the degree of the root, $k_{\max} \sim 2^n = O(N_n^{1/2})$, indicating that the nonzero fractal exponent in this regime reflects the extremely large degree of the root rather than criticality.
Indeed, the expected number of nodes connected to the root at distance $r$ decays exponentially with $r$ for $p<\pc$ (not shown, but similar to Fig.~\ref{fig:distance}(a)), demonstrating that the root cluster is spatially localized.
Thus, despite the nonzero value of $\psi(p)$, the root cluster does not spread over long distances, and the system remains in a nonpercolating phase.

Figure~\ref{fig:distance}(a) shows the correlation function $C(r)$, defined as the probability that a node located at distance $r$ from the root vertex $v_{\rm A}$ belongs to the same cluster as the root.
For $p<\pc$, $C(r)$ decays exponentially with distance, indicating a finite correlation length.
By contrast, at $p=\pc$ and for $p>\pc$, the data exhibit a power-law decay over a broad range of distances, which is consistent with the form $C(r)\sim r^{-2[1-\psi(p)]}$~\footnote{
A heuristic way to derive this relation is to focus on the connection between the root and the degree-2 nodes, which constitute the majority of the nodes in $G_n$. 
On the DHL, the number of degree-2 nodes at each odd distance from the root $v_{\rm A}$ is always exactly $2^{n-1}$, which scales as $\sqrt{N_n}$. 
Assuming a power-law decay of the correlation function, $C(r)\sim r^{-\eta}$, the expected number of degree-2 nodes belonging to the root cluster within distance $r$ scales as
\[\sum_{\substack{r'\le r\\ r' \text{ odd}}} C(r')\,\sqrt{N_n}.\]
Taking $r$ up to the maximal distance from $v_{\rm A}$, $L_n \sim \sqrt{N_n}$, we obtain
\[\sqrt{N_n} \sum_{r' \lesssim L_n} (r')^{-\eta} \sim N_n^{1/2} \, N_n^{(1-\eta)/2} = N_n^{1-\eta/2}.\]
Since the degree-2 nodes give the dominant contribution to the root-cluster mass, this quantity is expected to scale in the same way as the mean root-cluster size, $N_n^\psi$. 
This leads to $\psi=1-\eta/2$, or equivalently, $\eta=2(1-\psi)$.
}, reflecting the criticality of the phase.
This distinction is further illustrated in Fig.~\ref{fig:distance}(b), which shows the cumulative sum of the correlation function, $\sum_{r' \le r} C(r')$.
If $C(r)$ decays sufficiently fast, this cumulative sum saturates at large $r$, as indeed observed for $p<\pc$.
At $p=\pc$ and throughout the region $p>\pc$, however, the cumulative sum continues to grow without any sign of saturation up to the system size.
This behavior indicates that the correlation length remains divergent at $p=\pc$ and throughout the critical phase for $p>\pc$.

The persistence of criticality is also evident in the behavior of the susceptibility $\chi_n(p)$, which is defined as the mean size of all clusters except those containing the roots.
Figure~\ref{fig:susceptibility} shows $\chi_n(p)$ as a function of $p$.
As $p$ increases toward $\pc$ from below, $\chi_n(p)$ grows rapidly.
The susceptibility $\chi_n(p)$ diverges at the critical point and remains divergent throughout the entire region $p>\pc$, which is characteristic of a critical phase.

The behavior of the susceptibility shown in Fig.~\ref{fig:susceptibility} is also reminiscent of BKT transitions.
This similarity becomes more evident by expanding the recursion relation~\eqref{eq:recursion_Pn_site} for the conditional contact probability near the critical fixed point $(\pc, P_\ast(\pc))$.
Letting
\be
p = \pc -\epsilon, \quad P_n = P_\ast(\pc) + y_n,
\ee
and expanding in $\epsilon$ and $y_n$ up to second order, we obtain
\be
y_{n+1} - y_n= - a \epsilon - b y_n^2 + O(\epsilon^2, y_n^3), \label{eq:yn_rec}
\ee
with $a = 128/243$ and $b = 27/16$.
When $\epsilon$ is small, $y_n=0$ behaves as an almost stable fixed point of the RG recursion relation, and the recursion requires a very large number of RG steps $n$ to escape from its vicinity.
Let $n_\ast(\epsilon)$ denote the number of RG steps at which $y_n$ becomes $-O(1)$.
For $n \ll n_\ast$, $G_n$ exhibits critical behavior similar to that at $p=\pc$, whereas for $n \gg n_\ast$, it behaves as in the nonpercolating phase.
Therefore, the characteristic length scale $L_{n_\ast}$ associated with $G_{n_\ast}$ can be regarded as the correlation length in the present system.
The quantity $n_\ast$ can be estimated as follows.
Approximating Eq.~\eqref{eq:yn_rec} by a differential equation near the critical point, we obtain
\be
n_\ast \approx \int_{\Lambda}^{-\Lambda}\frac{dn}{dy}dy = \int_{-\Lambda}^{\Lambda}\frac{dy}{a\epsilon + b y^2}
\approx \int_{-\infty}^{\infty}\frac{dy}{a\epsilon + b y^2}= \frac{\pi}{\sqrt{ab}} \epsilon^{-1/2} \propto (\pc - p)^{-1/2}.
\ee
Here, $\Lambda=O(1)$ is a positive constant independent of $\epsilon$.
Since the root-to-root distance scales as $L_n \sim 2^n$, the correlation length can be regarded as
\be
\xi(p) \sim L_{n_\ast} \sim 2^{n_\ast}
\sim \exp\!\left( \frac{c}{\sqrt{\pc - p}} \right), \label{eq:corrLength}
\ee
where $c = \pi/\sqrt{a b}$, demonstrating an essential singularity at the critical point.

We assume the following finite-size scaling forms for $P_n(p)$, $\bsrt_n(p)$, and $\chi_n(p)$:
\be
P_n(p) = f_P\left(\frac{L_n}{\xi(p)}\right), \quad
\bsrt_n(p) = N_n^{\psi(\pc)} f_s\left(\frac{L_n}{\xi(p)}\right), \quad
\chi_n(p) = N_n^{2 \psi(\pc)-1} f_\chi\left(\frac{L_n}{\xi(p)}\right),
\ee
where the scaling functions $f_P(x)$, $f_s(x)$, and $f_\chi(x)$ approach constants for $x \ll 1$ and decay rapidly for $x \gg 1$.
Figure~\ref{fig:scaling} shows the finite-size scaling results using the correlation length $\xi(p)$ given by Eq.~\eqref{eq:corrLength}.
All data collapse onto universal curves, providing strong evidence that the correlation length diverges with a BKT-type essential singularity.

\section{Summary and Discussion}

In this work, we investigated the phase transition and critical phenomena in site percolation on the DHL by an exact generating-function analysis.
Our analysis revealed that site percolation on the DHL exhibits no percolating phase at any occupation probability $p<1$.
Instead, the system displays a nonpercolating phase for $p<\pc$ and a critical phase for $p>\pc$.
In the latter regime, the system never develops a giant cluster occupying a finite fraction of nodes, but is composed of subextensive clusters whose size scales as $N_n^{\psi(p)}$, where the fractal exponent $\psi(p)$ varies continuously with $p$.
Moreover, the susceptibility remains divergent throughout this phase, and the cluster-size distribution follows a power-law form. 
These features distinguish the regime $p>\pc$ from a conventional nonpercolating phase despite the absence of a giant cluster.

The transition at $\pc$ is governed by an essential singularity.
By expanding the recursion relation for the contact probability around the critical point, we showed that the correlation length diverges as $\xi(p) \sim \exp ({\rm const}/\sqrt{\pc-p})$.
Finite-size scaling analyses of the contact probability, the mean root-cluster size, and the susceptibility further showed excellent data collapse when these quantities are plotted against $L_n/\xi(p)$. 
In this restricted sense, the transition is of BKT-type, although the underlying mechanism is entirely different from that in the two-dimensional XY model.

This result is particularly striking in light of previous studies of critical phases in percolation.
As exemplified by nonamenable graphs, critical phases are generally understood to arise from a separation between the correlation length and correlation volume: the former remains finite, whereas the latter diverges.
This scenario relies on exponential volume growth, or equivalently, the small-world property, where the number of nodes within distance $\ell$ grows exponentially with $\ell$. 
In such systems, a cluster starting from a node can extend to arbitrarily large distances even when the correlation length remains finite.
The DHL, however, is finite-dimensional and does not possess the small-world property.
The emergence of a critical phase on the DHL therefore demonstrates that exponential volume growth is not a necessary condition for the emergence of a critical phase in percolation.
More importantly, the present system shows that a diverging correlation length does not necessarily lead to the formation of a giant cluster.

We argue that site dilution (node removal) on the hierarchical lattice qualitatively alters the renormalization behavior relative to that on Euclidean lattices, leading to the emergence of a critical phase.
Consider $p=1-\epsilon$ with $\epsilon\ll1$, so that only a small fraction of nodes is removed.
On Euclidean lattices, such sparse dilution acts as a short-range perturbation and becomes negligible under coarse graining.
On the DHL, by contrast, when the root nodes are removed with a nonzero probability, their effect remains relevant under coarse graining.
Node removal on the DHL creates connectivity bottlenecks at every hierarchical level, causing the effect of dilution to persist self-similarly across scales.
Even though clusters connected to a given node can extend over arbitrarily long distances, these bottlenecks prevent such clusters from coalescing into a single giant cluster, thereby giving rise to a critical phase.

This mechanism is specific to node removal and does not apply to edge removal.
On the same DHL, bond percolation exhibits a conventional percolating phase, in which the divergence of the correlation length leads directly to the emergence of a giant cluster.
This sharp contrast reflects a fundamental asymmetry under renormalization: node removal creates connectivity bottlenecks that persist across hierarchical levels, whereas the effects of edge removal remain short-ranged and become irrelevant under coarse graining.

The present work also provides a new perspective on the previous result on site-bond percolation in the decorated (2,2)-flower~\cite{hasegawa2012phase}.
In that system, which has the small-world property, bond percolation by itself gives rise to a critical phase below the percolating phase, whereas site dilution suppresses the emergence of a giant cluster and leads to a qualitatively different critical phase, thereby producing a transition between two critical phases.
From the present viewpoint, these two critical phases may originate from different dominant mechanisms: one associated with exponential volume growth and finite correlation length, and the other with scale-invariant connectivity bottlenecks that suppress global ordering despite a diverging correlation length.

More broadly, the present study shows that critical phases in percolation can arise even in finite-dimensional fractal networks, provided that hierarchical organization and node removal combine so that the effect of dilution survives across scales.
This finding extends the conventional understanding of the conditions under which critical phases arise.
An important direction for future work is to clarify how broadly this mechanism applies to other hierarchical fractal networks, including more general generator-based constructions~\cite{yakubo2022general}, as well as to stochastic networks, since the effects of fractal attributes on percolation thresholds and critical behavior remain to be clarified~\cite{balankin2018percolation, cruz2023percolation}.
It will also be important to investigate the implications of this mechanism for dynamical processes involving node failure, such as epidemic spreading and cascading failures.

\section*{Acknowledgment}

This work was supported by JSPS KAKENHI (24K06879) and JST ERATO SAKAI Real and Abstract Gels Project (JPMJER2401).


\appendix

\section{Recursion Relations for Derived Quantities}\label{sec:recForms}

In this appendix, we summarize the recursion relations used to explicitly evaluate the mean root-cluster size, the cluster-size distribution, and the susceptibility in site percolation on the diamond hierarchical lattice (DHL).

\subsection{Mean root-cluster size}

We define the first derivatives of the generating functions \eqref{eq:defT}--\eqref{eq:defV} at $x=1$ as
\be
\bt_n = \frac{d}{dx}T_n(x) \Big|_{x=1}, \quad
\bw_n = \frac{d}{dx}W_n(x,1) \Big|_{x=1} = \frac{d}{dx}W_n(1,x) \Big|_{x=1} =\frac{1}{2}\frac{d}{dx}W_n(x,x) \Big|_{x=1}, \quad
\bv_n = \frac{d}{dx}V_n(x) \Big|_{x=1} . \nonumber
\ee
By differentiating Eqs.~\eqref{eq:Trec}--\eqref{eq:Vrec} with respect to $x$ and evaluating them at $x=1$, we obtain the recursion relations for $\bt_n$, $\bw_n$, and $\bv_n$:
\ba
\bt_{n+1} &=& 
2p P_n^2 (1+2p P_nQ_n) + 4p P_n (1+ p P_nQ_n) \bt_n +4p^2P_n^2(1+P_n) \bw_n +4pq P_n^2 \bv_n, \label{eq:bt}
\\
\bw_{n+1} &=& 
2pP_nQ_n(1-pP_n^2) + 2pQ_n(1-pP_n^2)\bt_n +2p(1+P_n)(1-pP_n^2)\bw_n +2q(1-pP_n^2)\bv_n, \label{eq:bw}
\\
\bv_{n+1} &=& 2pP_n + 2p \bt_n + 2p \bw_n + 2(1-p Q_n) \bv_n, \label{eq:bv}
\ea
with the initial conditions
\be
\bt_1=0,\quad \bw_1=0,\quad \bv_1=0.
\ee
By iterating Eqs.~\eqref{eq:bt}--\eqref{eq:bv} and substituting the results into Eq.~\eqref{eq:sroot}, we obtain the mean root-cluster size $\bsrt_n(p)$ on $G_n$.

\subsection{Cluster-size distribution}

The cluster-size distribution $\ns_n(s)$ is obtained from the generating function $U_n(x)$ defined in Eq.~\eqref{eq:defU}.
To evaluate $U_n(x)$, we decompose it according to the states of the two roots as
\be
U_n(x) = p^2 \upp_n(x) + 2pq \upq_n(x) + q^2 \uqq_n(x).
\ee
Here, $\upp_n(x)$ is the generating function for the number of clusters of size $s$ under the condition that both roots are occupied, $\upq_n(x)$ corresponds to the case in which one root is occupied and the other is unoccupied, and $\uqq_n(x)$ corresponds to the case in which both roots are unoccupied.
By exploiting the recursive structure of the DHL, we obtain
\ba
\upp_{n+1}(x)
&=& 2 p^2 [2\upp_n(x) + xW_n^2(x,1)]  + 2 q^2 [2\upq_n(x)+1]\nonumber \\
&& + 2pq [2\upp_n(x)+2\upq_n(x)+xW_n^2(x,1) +1],
\\
\upq_{n+1}(x) 
&=& 2 p^2 [\upp_n(x) + \upq_n(x) + xW_n(x,1)V_n(x)] + 2 q^2 [\upq_n(x) + \uqq_n(x) + 1]\nonumber \\
&& + 2pq[\upp_n(x) + 2\upq_n(x) + \uqq_n(x) + 1 + xW_n(x,1)V_n(x)],
\\
\uqq_{n+1}(x) 
&=& 2 p^2 [2 \upq_n(x) + x V_n^2(x)] + 2 q^2 [2\uqq_n(x)+1] \nonumber \\
&&+2pq [2\upq_n(x)+2\uqq_n(x)+1+xV_n^2(x)].
\ea
Here, $V_n(x)$ and $W_n(x,1)$ obey the recursion relations given in Eqs.~\eqref{eq:Trec}--\eqref{eq:Vrec}.
From these equations, the coefficients of the polynomial $U_n(x)$ can be evaluated, and the cluster-size distribution $\ns_n(s)$ on $G_n$ is obtained via Eq.~\eqref{eq:nsFromUn}.

The full cluster-size distribution $\tilde{\mathcal{P}}_n(s)$ is obtained by adding the contribution of the clusters containing the roots.
Let $\tilde{U}_n(x)$ denote the generating function for the number of clusters of size $s$ on $G_n$ including the clusters containing the roots.
Then,
\be
\tilde{U}_n(x) = U_n(x) + p^2 [x^2 T_n(x) + 2xW_n(x,1)] + 2pq [xV_n(x) + 1] + 2 q^2,
\ee
and the corresponding cluster-size distribution is given by
\be
\tilde{\mathcal{P}}_n(s) = \frac{1}{N_n} \frac{1}{s!}\frac{d^s}{dx^s} \tilde{U}_n(x) \Big|_{x=0}.
\ee

\subsection{Susceptibility}

The susceptibility is given by
\be
\chi_n = \sum_s s^2 \ns_n(s) = \frac{1}{N_n} [U_n''(1) + U_n'(1)]. \nonumber
\ee
We decompose $U_n'(1)$ and $U_n''(1)$  as
\be
U_n'(1) = p^2 \bupp_n + 2pq \bupq_n + q^2 \buqq_n, \quad {\rm and} \quad
U_n''(1) = p^2 \buupp_n + 2pq \buupq_n + q^2 \buuqq_n,
\ee
respectively.
Here,
\be
\bupp_n = \frac{d}{dx} \upp_n(x) \Big|_{x=1}, \quad
\bupq_n = \frac{d}{dx} \upq_n(x) \Big|_{x=1}, \quad
\buqq_n = \frac{d}{dx} \uqq_n(x) \Big|_{x=1},
\ee
and
\be
\buupp_n = \frac{d^2}{dx^2} \upp_n(x) \Big|_{x=1}, \quad
\buupq_n = \frac{d^2}{dx^2} \upq_n(x) \Big|_{x=1}, \quad
\buuqq_n = \frac{d^2}{dx^2} \uqq_n(x) \Big|_{x=1}.
\ee

The recursion relations for $\bupp_n$, $\bupq_n$, and $\buqq_n$ are given by
\ba
\bupp_{n+1} &=& 2 p Q_n^2 + 4 p Q_n \bw_n +4 p\bupp_n + 4q \bupq_n,
\\
\bupq_{n+1} &=& 2pQ_n +2p\bw_n + 2pQ_n \bv_n + 2p\bupp_n +2\bupq_n + 2q\buqq_n,
\\
\buqq_{n+1} &=& 2p + 4p\bv_n + 4p\bupq_n + 4q\buqq_n,
\ea
with the initial conditions $\bupp_1 = \bupq_1 = \buqq_1 =0$.

The recursion relations for $\buupp_n$, $\buupq_n$, and $\buuqq_n$ are given by
\ba
\buupp_{n+1} &=& 4p\buupp_n + 4q\buupq_n + 4pQ_n\btw_n + 8pQ_n \bw_n + 4p\bw_n^2, \label{eq:buupp}
\\
\buupq_{n+1} &=& 2p\buupp_n + 2\buupq_n + 2q\buuqq_n + 2pQ_n\btv_n + 4pQ_n \bv_n + 2p\btw_n + 4p\bw_n + 4p\bv_n \bw_n, \label{eq:buupq}
\\
\buuqq_{n+1} &=& 4p\buupq_n + 4q\buuqq_n + 4p\btv_n + 8p\bv_n + 4p\bv_n^2. \label{eq:buuqq}
\ea
The initial conditions are $\buupp_1 = \buupq_1 = \buuqq_1 =0$.

The evaluation of Eqs.~\eqref{eq:buupp}--\eqref{eq:buuqq} also requires the following quantities:
\be
\begin{split}
\btt_n &= \frac{d^2}{dx^2} T_n(x) \Big|_{x=1}, \quad
\btw_n=\frac{d^2}{dx^2} W_n(x,1) \Big|_{x=1}, \\
\bsw_n &=\frac{d}{dy}\frac{d}{dx} W_n(x, y) \Big|_{x=1, y=1}, \quad
\btv_n=\frac{d^2}{dx^2} V_n(x) \Big|_{x=1}.
\end{split}
\ee
The recursion relations for $\btt_n$, $\bsw_n$, $\btw_n$, and $\btv_n$ are given by
\ba
\btt_{n+1} &=& 2 p^2 P_n^3 (1+3P_n) + 4pP_n (1+pQ_n) \btt_n + 4pqP_n^2\btv_n + 4p^2P_n^2(1+P_n)\btw_n \nonumber \\
&&+ 8p^2P_n^3\bsw_n + 8pP_n[1+p(1+3P_n)P_n]\bt_n + 4p[1+2p(1+Q_n)P_n]\bt_n^2 + 8pqP_n^2\bv_n \\
&& + 4pqP_n^2\bv_n^2 + 8p^2P_n^2(1+3P_n)\bw_n + 4p^2P_n^2\bw_n^2 + 16pqP_n\bt_n\bv_n + 16p^2P_n(1+2P_n)\bt_n \bw_n, \nonumber 
\\
\bsw_{n+1} &=& 2p^2P_n^2Q_n^2 + 4p^2P_nQ_n^2\bt_n + 2p^2Q_n^2\bt_n^2 + 2q(1+q-pP_n^2)\bv_n^2 + 4pqP_nQ_n\bv_n \nonumber \\
&&+ 2p(1+p+2pP_n)\bw_n^2 + 4pP_n(1+p-2pP_n^2)\bw_n + 4pP_n(1-pP_n^2) \bsw_n  \\
&&+ 4pqQ_n\bt_n\bv_n + 4p(1+p-2pP_n^2)\bt_n\bw_n + 4pq(1+P_n)\bv_n\bw_n, \nonumber
\\
\btw_{n+1} &=& 2p^2P_n^2Q_n^2 + 2pQ_n(1-pP_n^2)\btt_n +2q(1-pP_n^2)\btv_n +2p(1+P_n)(1-pP_n^2)\btw_n \nonumber \\
&&+ 4pQ_n[1+pP_n(1-2P_n)]\bt_n + 2p^2Q_n^2 \bt_n^2 + 4pqP_nQ_n\bv_n +2q^2\bv_n^2 + 4pP_n(1+p-2pP_n^2)\bw_n  \\
&&+ 2p^2(1+P_n)^2\bw_n^2 + 4pqQ_n\bt_n\bv_n + 4p(1+p-2pP_n^2)\bt_n\bw_n + 4pq(1+P_n)\bv_n\bw_n, \nonumber
\\
\btv_{n+1} &=& 2p^2P_n^2 + 2p\btt_n + 2(1-pQ_n)\btv_n + 2p\btw_n +4p(1+pP_n)\bt_n + 2p^2\bt_n^2 \nonumber \\
&&+ 4p(2-pQ_n)P_n\bv_n + 2(1-pQ_n)^2\bv_n^2 + 4p^2P_n\bw_n +2p^2\bw_n^2 \\
&&+ 4p(2-pQ_n)\bt_n\bv_n +4p^2\bt_n\bw_n +4p(1-pQ_n)\bv_n\bw_n. \nonumber
\ea

\section{Derivation of the Square-Root Singularity of the Fixed Point}\label{sec:square}

As shown in the main text, the fixed point $P \equiv P_\ast(p)$ of the conditional contact probability satisfies Eq.~\eqref{eq:recursion_Pn_site}.
We introduce
\be
F(p, P)=P-2pP^2+p^2P^4,
\ee
so that the fixed point is determined by $F(p, P)=0$.
Let us set
\be
p=\pc+\delta \ (\delta>0), \quad P=P_{\rm c}+\Delta,
\ee
and expand $F(p, P)$ around $(\pc, P_{\rm c})$, where $P_{\rm c}=P_\ast(\pc)$.
Since $F(\pc, P_{\rm c})=0$ and $F_P(\pc, P_{\rm c})=0$, the Taylor expansion yields
\begin{equation}
F(p, P) = F_p(\pc, P_{\rm c}) \delta +\frac{1}{2}F_{PP}(\pc, P_{\rm c}) \Delta^2 +O(\delta^{3/2}).
\label{eq:balance}
\end{equation}
Here, higher-order terms such as $\delta\Delta$ and $\delta^2$ are subleading, consistent with the scaling $\Delta=O(\delta^{1/2})$ obtained below.
The relevant derivatives evaluated at $(\pc, P_{\rm c})$ are $F_p(\pc, P_{\rm c})=-128/243$ and $F_{PP}(\pc, P_{\rm c})=27/8$.
Substituting these into Eq.~\eqref{eq:balance}, we obtain
\begin{equation}
-\frac{128}{243} \delta +\frac{1}{2}\frac{27}{8} \Delta^2 =0,
\end{equation}
which leads to
\begin{equation}
\Delta^2=\frac{2048}{6561} \delta + O(\delta^{3/2}). \label{eq:Deltadelta}
\end{equation}
Therefore, for $p>\pc$,
\begin{equation}
P_\ast(p)-P_\ast(\pc) = \Delta \sim \sqrt{p-\pc}. 
\end{equation}
This establishes the square-root singularity \eqref{eq:squareRoot} of the fixed point at the critical point.

\section{Cluster-size Distribution}\label{sec:ns}

\begin{figure}[t]
\centering
(a)
\includegraphics[width=0.45\linewidth]{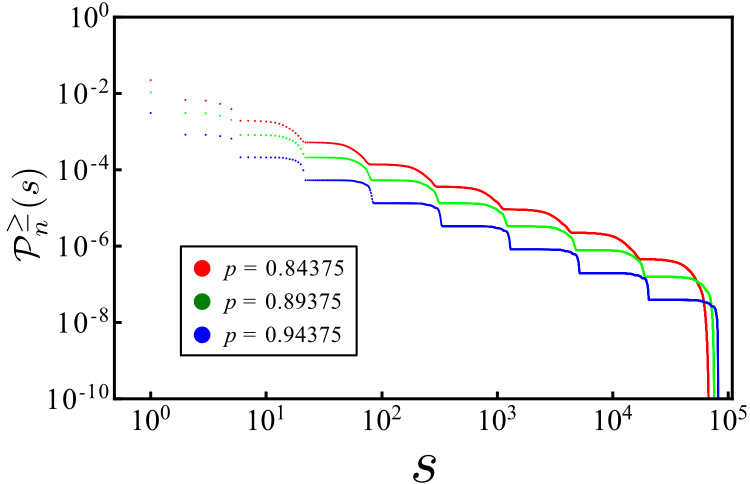}
(b)
\includegraphics[width=0.45\linewidth]{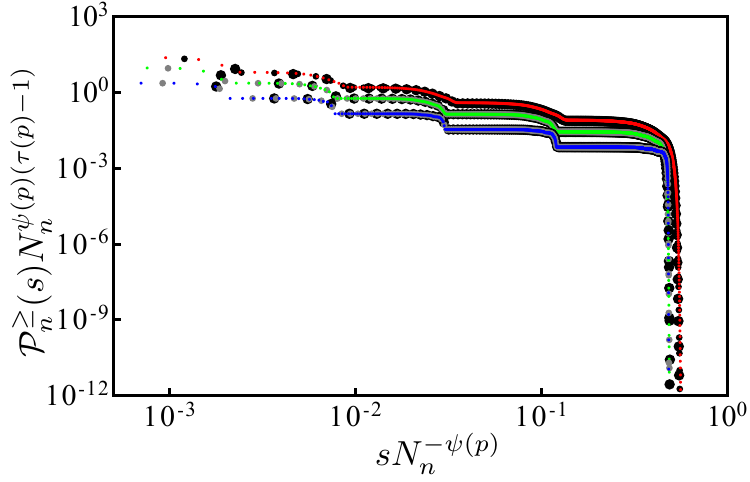}
\caption{
Cumulative cluster-size distribution $\ns_n^{\geq}(s)$ for site percolation on the DHL.
(a) $\ns_n^{\geq}(s)$ for generation $n=10$ at $p=0.84375 (=\pc)$ (red), $p=0.89375$ (green), and $p=0.94375$ (blue).
(b) Finite-size scaling plots of $\ns_n^{\geq}(s)$ for the same values of $p$. 
Data for generations $n=8, 9$, and $10$ are shown: 
large black symbols correspond to $n=8$, medium gray symbols to $n=9$, and small colored symbols to $n=10$ (red: $p=0.84375$, green: $p=0.89375$, blue: $p=0.94375$).
The collapse of the data for different generations supports the finite-size scaling form of the cumulative cluster-size distribution in the critical phase.
}
\label{fig:ns}
\end{figure}

We present the cluster-size distribution for site percolation on the DHL.
We focus on the cumulative cluster-size distribution, $\ns_n^{\geq}(s) = \sum_{s^\prime \geq s} \ns_n(s^\prime)$, which provides a clearer characterization of the tail behavior.
Figure~\ref{fig:ns}(a) shows $\ns_n^{\geq}(s)$ for generation $n=10$ at the critical point $p=\pc$ and in the critical phase for $p>\pc$, obtained from the generating function method. 
We note that the cumulative distributions computed using the generating functions are in good agreement with those obtained from Monte Carlo simulations (not shown).
At $p=\pc$, the cumulative distribution exhibits a power-law-like decay over several decades of cluster size.
For $p>\pc$ as well, the distribution retains this form, with the overall scale and effective slope depending continuously on the occupation probability $p$.
However, the slope is shallow, and its variation with $p$ is correspondingly slow.

We carry out finite-size scaling of the cumulative cluster-size distribution $\ns_n^{\geq}(s)$ for $p>\pc$, and confirm that it follows a power-law form, with the exponent $\tau$ varying with $p$.  
We assume the following finite-size scaling form~\cite{nogawa2009monte} for $\ns_n^{\geq}(s)$:
\begin{equation}
\ns_n^{\geq}(s) = N_n^{-\psi(p)(\tau(p)-1)}f\left(s N_n^{-\psi(p)}\right),
\end{equation}  
where the scaling function $f(x)$ behaves as  
\begin{equation}
f(x) \sim 
\begin{cases}
\text{rapidly decaying function} & \text{for} \quad x \gg 1, \\
x^{1 - \tau(p)} & \text{for} \quad x \ll 1,
\end{cases}
\end{equation}
and the exponent $\tau(p)$ is related to the fractal exponent $\psi(p)$ as  $\tau(p) = 1 + \psi^{-1}(p)$.
Figure~\ref{fig:ns}(b) shows the scaling results for $\ns_n^{\geq}(s)$ for $n=8,9,10$.
The data collapse onto a single curve for each value of $p$, confirming the validity of the scaling form.  
This implies that, in the large size limit, the cluster-size distribution $\ns_n(s)$ (as well as the full cluster-size distribution $\tilde{\ns}_n(s)$) follows a power law of the form $\ns_n(s) \propto s^{-\tau(p)}$ in the critical phase.



\end{document}